\documentclass[a4paper, 12pt]{article}
\usepackage{epsfig}
\usepackage{amsmath}
\usepackage{amssymb}
\usepackage[authoryear]{natbib}

\usepackage[amsmath,thmmarks]{ntheorem}

\theorembodyfont{\slshape}
\newtheorem{Thm}{Theorem}[section]

\newtheorem{Lem}{Lemma}[section]
\newtheorem{Propn}{Proposition}[section]
\newtheorem{Cor}{Corollary}[section]

\usepackage{setspace}
\usepackage{algorithm,algorithmic}

\newcommand{\T}{\textstyle}

\newcommand{\inv}{^{-1}}
\newcommand{\N}{\mathcal{N}}

\newcommand{\hopt}{h_{\text{opt}}}

\newcommand{\simiid}{\stackrel{\text{iid}}{\sim}}

\newcommand{\rmd}{\mathrm{d}}
\newcommand{\mb}{\boldsymbol}

\newcommand{\E}{\mathbb{E}}
\newcommand{\V}{\mathbb{V}}
\newcommand{\real}{\mathbb{R}}
\newcommand{\cov}{\mathrm{cov}}

\newcommand{\prob}{\mathbb{P}}
\newcommand{\I}{\mathbb{I}}
\renewcommand{\qed}{\hfill\ensuremath{\Box}}

\begin{document}

\title{An Adaptive Sequential Monte Carlo Sampler}
\author{Paul Fearnhead and Benjamin M. Taylor}
\maketitle

\begin{abstract}
Sequential Monte Carlo (SMC) methods are not only a popular tool in the analysis of state--space models, but offer an alternative to MCMC in situations where Bayesian inference must proceed via simulation. This paper introduces a new SMC method that uses adaptive MCMC kernels for particle dynamics. The proposed algorithm features an online stochastic optimization procedure to select the best MCMC kernel and simultaneously learn optimal tuning parameters. Theoretical results are presented that justify the approach and give guidance on how it should be implemented. Empirical results, based on analysing data from mixture models, show that the new adaptive SMC algorithm (ASMC) can both choose the best MCMC kernel, and learn an appropriate scaling for it. ASMC with a choice between kernels outperformed the adaptive MCMC algorithm of \cite{haario1998} in 5 out of the 6 cases considered. 
\end{abstract}


\section{Introduction}

Sequential Monte Carlo (SMC) is a class of algorithms that enable simulation from a target distribution of interest. These algorithms are based on defining a series of distributions, and generating samples from each distribution in turn. SMC was initially used in the analysis of state-space models. In this setting there is a time--evolving hidden state of interest, inference about which is based on a set of noisy observations \citep{gordon1993,liuchen1998,SMCMiP,fearnhead2002}. The sequence of distributions are defined to be the posterior distributions of the state at consecutive time-points given the observations up to those time points. More recent work has looked at developing SMC methods that can analyse state-space models which have unknown fixed parameters. Such methods introduce steps into the algorithm to allow the support of the sample of parameter values to change over time, for example by using ideas from kernel density estimation \citep{SMCMPliuwest}, or MCMC moves \citep{gilks2001,storvik2002,fearnhead2002}.

Most recently, SMC methods have been applied as an alternative to MCMC for standard Bayesian inference problems. \citep{neal2001,chopin2002,delmoral2006,fearnhead2008}. In this paper the focus will be on methods for sampling from the posterior distribution of a set of parameters of interest. SMC methods for this class of targets introduce an artificial sequence of distributions that run from the prior to the posterior and sample recursively from these using a combination of Importance Sampling and MCMC moves. This approach to sampling has been demonstrated empirically to often be more effective than using a single MCMC chain \citep{jasra2007,jasra2008}. There are heuristic reasons for why this may true in general: the annealing of the target and spread of samples over the support means that SMC is less likely to be become trapped in posterior modes.

Simply invoking an untuned MCMC move within an SMC algorithm would likely lead to poor results because the move step would not be effective in combating sample depletion. The structure of SMC means that at the time of a move there is a sample from the target readily available, this can be used to compute posterior moments and inform the shape of the proposal kernel as in \cite{jasra2008a}; however, further refinements can lead to even better performance. Such refinements include the scaling of estimated target moments by an optimal factor, see \cite{roberts2001} for example. For general targets and proposals no theoretical results for the choice of scaling exist, and this has led to the recent popularity of adaptive MCMC \citep{haario1998,andrieu2001,roberts2009,craiu2009,andrieu2008}. In this paper the idea of adapting the MCMC kernel within an SMC algorithm will be explored.

To date there has been little work at adapting SMC methods. Exceptions include the method of \cite{jasra2008a}, whose method assumes a likelihood tempered sequence of target densities (see \cite{neal2001}) and the adaptation procedure both chooses this sequence online, as well as computing the variance of a random walk proposal kernel used for particle dynamics. \cite{cornebise2008} also considers adapting the proposal distribution within SMC for state-space models. Assuming that the proposal density belongs to a parametric family with parameter $\theta$, their method proceeds by simulating a number of realisations for each of a range of values of $\theta$ and selecting the value that minimises the empirical Shannon entropy of the importance weights; new samples are then re--proposed using this approximately optimal value. Further related work includes that of \cite{douc2005a} and \cite{cappe2008} on respectively population Monte Carlo and adaptive importance sampling.

The aims of this paper are to introduce a new adaptive SMC algorithm (ASMC) that automatically tunes MCMC move kernels and chooses between different proposal densities and to provide theoretical justification of the method. The algorithm is based on having a distribution of kernels and their tuning parameters at each iteration. Each current sample value, called a particle, is moved using an MCMC kernel drawn from this distribution. By observing the expected square jumping distance \citep{craiu2009,sherlock2009} for each particle it is possible to learn which MCMC kernels are mixing better. The information thus obtained can then used to update the distribution of kernels. The key assumption of the new approach is that the optimal MCMC kernel for moving particles does not change much over the iterations of the SMC algorithm. As will be discussed, and shown empirically, in section \ref{sect:resultsASMC} this can often be achieved by appropriate parameterisation of a family of MCMC kernels.

The structure of the paper is as follows. In the next section, the model of interest will be introduced and followed by a review of MCMC and SMC approaches. Then in Section \ref{sect:newmethod}, the new adaptive SMC will be presented. Guidelines on implementing the algorithm as well as some theory on the convergence will be presented in Section \ref{sect:theoryASMC}. In Section \ref{sect:resultsASMC} the method will be evaluated using simulated data. The results show that the proposed method can successfully choose both an appropriate MCMC kernel and an appropriate scaling for the kernel. The paper ends with a discussion. 

\section{Model}

The focus of this paper will be on Bayesian inference for parameters, $\theta$, from a model where independent identically distributed data is available. Note that the ideas behind the proposed adaptive SMC algorithm can be applied more generally (see section \ref{sect:discuss_extASMC}). Let $\pi(\theta)$ denote the prior for $\theta$ and $\pi(y|\theta)$ the probability density for the observations. The aim will be to calculate the posterior density,
\begin{equation} \label{eq:1}
   \pi(\theta|y_{1:n})\propto\pi(\theta)\prod_{i=1}^n\pi(y_i|\theta),
\end{equation}
where, here and throughout, $\pi$ will be used to denote a probability density, and $y_{1:t}$ means $y_1,\ldots,y_t$. 

In general, $\pi(\theta|y_{1:n})$ is analytically intractable and so to compute posterior functionals of interest, for example expectations, Monte Carlo simulation methods are often employed. Sections \ref{sect:MCMCintro} and \ref{sect:SMCintro} provide a brief description of two such Monte Carlo approaches.

\subsection{MCMC\label{sect:MCMCintro}}

An MCMC transition kernel, $K_h$, is an iterative rule for generating samples from a target probability density, for example a posterior. $K_h$ comprises a proposal kernel, here and throughout denoted $q_h$ (the subscript $h$ indicates dependence on a tuning parameter) and an acceptance ratio that depends on the target and, in general, the proposal densities (see \cite{MCMCiP,gamermanlopes} for reviews of MCMC methodology). The most generally applicable MCMC method is Metropolis--Hastings, see Algorithm \ref{alg:met_hast} \citep{metropolis1953,hastings1970}.

\begin{algorithm}
   \caption{Metropolis--Hastings Algorithm \citep{metropolis1953,hastings1970}}
   \label{alg:met_hast}
   \begin{algorithmic}[1]
      \STATE Start with an initial sample, $\theta^{(0)}$, drawn from any density, $\pi_0$.
      \FOR{$j=1,2,\ldots$}
         \STATE Propose a move to a new location, $\tilde\theta$, by drawing a sample from $q_h(\theta^{(i-1)},\tilde\theta)$.
         \STATE Accept the move (ie set $\theta^{(i)} = \tilde\theta$) with probability,
         \begin{equation}\label{eqn:MHacc}
            \min\left\{1,\frac{\pi(\tilde\theta|y_{1:n})}{\pi(\theta^{(i-1)}|y_{1:n})}\frac{q_h(\tilde\theta,\theta^{(i-1)})}{q_h(\theta^{(i-1)},\tilde\theta)}\right\},
         \end{equation}
         else set $\theta^{(i)} = \theta^{(i-1)}$.
      \ENDFOR
   \end{algorithmic}
\end{algorithm}

Probably the simplest MH algorithm is the random walk Metropolis (RWM). The proposal kernel for RWM is a symmetric density centred on the current state, the most common example being a multivariate normal, $q_h(\theta^{(i-1)},\tilde\theta)=\N(\tilde\theta;\theta^{(i-1)},h^2\hat\Sigma_\pi)$, where $\hat\Sigma_\pi$ is an estimate of the target covariance. Both the values of $\hat\Sigma_\pi$ and $h$ are critical to the performance of the algorithm. If $\hat\Sigma_\pi$ does not accurately estimate the posterior covariance matrix, then the likely directions of the random walk moves will likely be inappropriate. On the other hand, a value of $h$ that is too small will lead to high acceptance rates, but the samples will be highly correlated. If $h$ is too large then the algorithm will rarely move, which in the worst case scenario could lead to a degenerate sample. 

These observations on the r\^ole of $h$ point to the idea of an \emph{optimal scaling}, a $h$ somewhere between the extremes that promotes the best mixing of the algorithm. In the case of elliptically symmetric unimodal targets, an optimal random walk scaling can sometimes be computed numerically; this class of targets includes the Multivariate Gaussian \citep{sherlock2009}. Other theoretical results include optimal acceptance rates which are derived in the limit as the dimension of $\theta$, $d\rightarrow\infty$ (see \cite{roberts2001} for examples of targets and proposals). In general however, there are no such theoretical results. 

One way of circumventing the need for analytical optimal scalings is to try to learn them online \citep{andrieu2001,atchade2005}, this can include learning both a good scaling, $h$, and estimating the target covariance, $\hat\Sigma_\pi$ \citep{haario1998}. Recent research in adaptive MCMC has generated a number of new algorithms (see for example \cite{andrieu2008,roberts2009,craiu2009}), though some care must be taken to ensure that the resulting chain has the correct ergodic distribution.

\subsection{Sequential Monte Carlo\label{sect:SMCintro}}

An alternative approach to generating samples from a posterior is to use sequential Monte Carlo (SMC, see \cite{delmoral2006} for a review). The main idea behind SMC is to introduce a sequence of densities leading from the prior to the target density of interest and to iteratively update an approximation to these densities. For the application considered here, it is natural to define these densities as $\pi_t(\theta)=\pi(\theta|y_{1:t})$ for $t=1,\ldots,n$; this `data tempered' schedule will be used in the sequel. The approximations to each density are defined in terms of a set of weighted particles, $\{\theta_t^{(j)},w_t^{(j)}\}_{j=1}^M$, produced so that as $M\rightarrow\infty$, Monte Carlo sums converge to their `correct' expectations:
\begin{equation*}\label{eqn:properweight}
   \lim_{M\rightarrow\infty} \left\{\frac{\sum_{j=1}^Mw_t^{(j)}f(\theta_t^{(j)})}{\sum_{i=1}^Mw_t^{(i)}}\right\} = \E_{\pi_t(\theta_t)}[f(\theta_t)],
\end{equation*}
for all $\pi_t$--integrable functions, $f$. One step of an SMC algorithm can involve importance reweighting, resampling and moving the particles via an MCMC kernel \citep{gilks2001,chopin2002}. For concreteness, this paper will focus on the iterated batch importance sampling (IBIS) algorithm of \cite{chopin2002}. 

The simplest way to update the particle approximation in model (\ref{eq:1}) is to let $\theta_{t}^{(j)}=\theta_{t-1}^{(j)}$ and
$w_t^{(j)}=w_{t-1}^{(j)}\pi(y_t|\theta_{t}^{(j)})$. However such an algorithm will degenerate for large $t$, as eventually only one particle will have non-negligible weight. Within IBIS, resample--move steps (sometimes referred to here as simply `move steps') are introduced to alleviate this. In a move step, the particles are first resampled so that the expected number of copies of particle $\theta_t^{(j)}$ is proportional to $w_t^{(j)}$. This process produces multiple copies of some particles. In order to create particle diversity, each resampled particle is moved by an MCMC kernel. The MCMC kernel is chosen to have stationary distribution $\pi_t$. The resulting particles are then assigned a weight of $1/M$.

The decision of whether to apply a resample-move step within IBIS is based on the effective sample size (ESS, see \cite{kong1994,liuchen1998}). The ESS is a measure of variability of the particle weights; using this to decide whether to resample is justified by arguments within \cite{liu1995} and \cite{liu1998}. Full details of IBIS are given in Algorithm \ref{alg:IBIS}.

\begin{algorithm}
   \caption{Chopin's IBIS algorithm}
   \label{alg:IBIS}
   \begin{algorithmic}[1]
      \STATE Initialise from the prior $\{\theta_0^{(j)},w_0^{(j)}\}_{j=1}^M\sim\pi_0$.
      \FOR{$t=1,\ldots,n$}
         \STATE Assume current $\{\theta_{t-1}^{(j)},w_{t-1}^{(j)}\}_{j=1}^M\sim\pi_{t-1}$
         \STATE Reweight $w_t^{(j)} = w_{t-1}^{(j)}\pi_t(\theta_{t-1}^{(j)})/\pi_{t-1}(\theta_{t-1}^{(j)})$. Result: $\{\theta_{t-1}^{(j)},w_t^{(j)}\}_{j=1}^M\sim\pi_t$.
         \IF{particle weights not degenerate (see text)}
            \STATE $\{\theta_{t}^{(j)},w_{t}^{(j)}\}_{j=1}^M \leftarrow \{\theta_{t-1}^{(j)},w_{t-1}^{(j)}\}_{j=1}^M$
            \STATE $t\rightarrow t+1$.
         \ELSE
            \STATE \label{alg:IBIS_state_resample} Resample: let $\mathcal{K}=\{k_1,\ldots,k_M\}\subseteq\{1,\ldots,M\}$ be the resampling indices, then $\{\theta_{t-1}^{(k)},1/M\}_{k\in\mathcal K}\sim\pi_t$. Relabel: $k_j \leftarrow j$, the $j$th resampling index so that $\{\theta_{t-1}^{(j)},1/M\}_{j=1}^M\sim\pi_t$.
            \STATE Move via $\pi_t$--invariant MCMC kernel. Result: $\{\theta_t^{(j)},1/M\}_{j=1}^M\sim\pi_t$.
         \ENDIF
      \ENDFOR
   \end{algorithmic}
\end{algorithm}

Chopin's IBIS algorithm is a special case of the resample--move (RM) algorithm of \cite{gilks2001} and the general algorithm described by \cite{delmoral2006} (note that the latter method applies beyond MCMC--within--SMC and provides a unifying framework for sampling from sequences of targets). The main difference between RM and IBIS is that, in their presentation of RM, \cite{gilks2001} use resampling and move steps at each iteration of the sampler. Chopin noticed that at a particular iteration it may be better to just reweight the particles, rather than incur the computational cost and degeneracy induced by a resample--move step. Another related algorithm, a development of simulated annealing \citep{kirkpatrick1983} due to \cite{neal2001}, utilises an alternative `likelihood tempered' sequence of targets. The proposed target sequence is $\pi_t(\theta)=\pi(\theta)\pi(y_{1:n}|\theta)^{\xi_t}$, where $\{\xi_t\}$ is a sequence of real numbers starting at $0$ (the prior) and ending on $1$ (the posterior). Since each move step requires evaluation of the likelihood over all available observations, the main disadvantage of likelihood tempering is computational cost, although for models with sufficient statistics this is not an issue. Further disadvantages of Neal's proposed algorithm are the absence of resampling steps which eventually leads to sample degeneracy; and the lack of interpretability of intermediate target densities.

The efficiency of an SMC algorithm, such as IBIS, depends on the mixing properties of the associated MCMC kernel. Within SMC there is the advantage of being able to use the current set of particles to help tune an MCMC kernel. For example, the weighted particles can give an estimate of the posterior covariance matrix, which can be used within a random walk proposal. However even in this case, the proposal variance still needs to be appropriately scaled \citep{roberts2001,sherlock2009}. In the next section the new adaptive SMC procedure will be introduced, the algorithm can learn an appropriate tuning for the MCMC kernel, and can also be used to choose between a set of possible kernels.

\section{The Adaptive SMC Sampler\label{sect:newmethod}} 

First consider the case where the move step in the IBIS algorithm involves one type of MCMC kernel. Let $\pi_t$ be an \emph{arbitrary} continuous probability density (the target) and $K_{h,t}$ a $\pi_t$--invariant MCMC kernel with tuning parameter, $h$. The parameter $h$ is to be chosen to maximise the following utility function,
\begin{eqnarray} \label{eqn:optfun}
   g^{(t)}(h) &=& \int \pi_t(\theta_{t-1})K_{h,t}(\theta_{t-1},\theta_t)\Lambda(\theta_{t-1},\theta_t)\rmd\theta_{t-1}\rmd\theta_t,\\
   &=& \E\left[\Lambda(\theta_{t-1},\theta_t)\right], \nonumber
\end{eqnarray}
where $\Lambda(\theta_{t-1},\theta_t)>0$ is a measure of mixing of the chain. Most MCMC adaptation criteria can be viewed in this way \citep{andrieu2008}. Note that for simplicity of presentation, $\Lambda$ only depends on the current and subsequent state, though the idea readily extends to more complex cost functionals, for example involving multiple transitions of the MCMC chain. The function $g^{(t)}$ is the average performance of the chain with respect to $\Lambda$, which would normally be some measure of mixing. The ideal choice for $\Lambda$ would be the integrated autocorrelation time (whence the goal would be to maximise $-g^{(t)}$), but a computationally simpler measure of mixing is the expected square jumping distance (ESJD). Maximising the ESJD is equivalent to minimising the lag-1 autocorrelation; this measure is often used within adaptive MCMC, see for example \cite{sherlock2009,pasarica2010}. 

In the following it will be assumed that the proposal distribution can depend on quantities calculated from the current set of particles (for example estimates of the posterior variance), but this will be suppressed in the notation.
The main idea of ASMC is to use the observed instances of $\Lambda(\theta_{t-1},\theta_t)$ to help choose the best $h$. The tuning parameter will be treated as an auxiliary random variable. At time-step $t$ the aim is to derive a density for the tunings, $\pi^{(t)}(h)$.  If a move step is invoked at this time, a sample of $M$ realisations from $\pi^{(t)}(h)$, denoted $\{h_t^{(j)}\}_{j=1}^M$, will be drawn and `allocated' to particles at random. 

When moving the $j$th resampled particle, the tuning parameter $h_t^{(j)}$ will be used within the proposal distribution.
For notational simplicity this value will be denoted $h$ in the following.
Let $\theta_{t-1}^{(j)}$ be the $j$th resampled particle (see step \ref{alg:IBIS_state_resample} of Algorithm \ref{alg:IBIS}). In moving this particle, $\tilde{\theta}_t^{(j)}$ is drawn from $q_{h}(\theta_{t-1}^{(j)},\,\cdot\,)$, and accepted with probability 
$\alpha_h(\theta_{t-1}^{(j)},\tilde\theta_t^{(j)})$, given by (\ref{eqn:MHacc}). If the proposed particle is accepted then $\theta_t^{(j)}=\tilde{\theta}_t^{(j)}$ otherwise $\theta_t^{(j)}={\theta}_{t-1}^{(j)}$.

The utility function in (\ref{eqn:optfun}) simplifies to,
\begin{equation*}
   g^{(t)}(h) = \int \pi_t(\theta_{t-1})q_h(\theta_{t-1},\tilde\theta_t)\tilde\Lambda(\theta_{t-1},\tilde\theta_t)\rmd\theta_{t-1}\rmd\tilde\theta_t,
\end{equation*}
where 
\begin{equation*}
   \tilde\Lambda(\theta_{t-1}^{(j)},\tilde\theta_t^{(j)})= \alpha_h(\theta_{t-1}^{(j)},\tilde\theta_t^{(j)})\Lambda(\theta_{t-1}^{(j)},\tilde\theta_t^{(j)}). 
\end{equation*}
Since by assumption the resampled particles are approximately drawn from $\pi_t$ and proposed particles are drawn from $q_h(\theta_{t-1},\tilde\theta_t)$, the quantity $\tilde\Lambda(\theta_{t-1}^{(j)},\tilde\theta_t^{(j)})$ can be viewed as an unbiased estimate of $g^{(t)}(h)$.

The approach in this paper is to use the observed $\tilde\Lambda(\theta_{t-1}^{(j)},\tilde\theta_t^{(j)})$ to update the distribution $\pi^{(t)}(h)$ to a new distribution $\pi^{(t+1)}(h)$. In particular each $h_t^{(j)}$ will be assigned a weight, $f(\tilde\Lambda(\theta_{t-1}^{(j)},\tilde\theta_t^{(j)}))$, for some function $f:\real^+\rightarrow\real^+$. The new density of scalings will be defined,
\begin{equation} \label{eq:pit}
 \pi^{(t+1)}(h) \propto \sum_{i=1}^M f(\tilde\Lambda(\theta_{t-1}^{(j)},\tilde\theta_t^{(j)})) R(h-h_t^{(j)}),
\end{equation}
where $R(h-h_t^{(j)})$ is a density for $h$ which is centred on $h_t^{(j)}$. Simulating from $\pi^{(t+1)}(h)$ is achieved by first resampling the $h_t^{(j)}$s with probabilities proportional to their weight and then adding noise to each resampled value; the distribution of this noise is given by $R(\,\cdot\,)$. The motivation for adding noise to the resampled $h$--values is to avoid the distributions $\pi^{(t)}(h)$ degenerating too quickly to a point-mass on a single value. Similar ideas are used in dynamic SMC methods for dealing with fixed parameters, for example \cite{west1993,SMCMPliuwest}. In practice the variance of the noise can depend on the variance of $\pi^{(t)}(h)$ and by analogy to Kernel density estimation should tend to 0 as the number of particles gets large.

If there is no resampling at step $t$ then set $\pi^{(t+1)}(h)=\pi^{(t)}(h)$. The scheme is initiated with an arbitrary distribution $\pi(h)$.
The specific choice of $f$ considered in this paper is a simple linear weighting scheme,
\begin{equation*}
   f(\tilde\Lambda) = a + \tilde\Lambda,\qquad a\geq 0.
\end{equation*} 
Theoretical justification for this choice is given in the next section.

One assumption of the proposed approach is that a good choice of $h$ at one time-step will be a good choice at nearby time-steps. Note that this is based on an implicit assumption within SMC that successive targets are similar (see \cite{chopin2002,delmoral2006} for example). Furthermore, using estimates of posterior variances within the proposal distribution can also help ensure that good values of $h$ at one time-step will be a good choice at nearby time-steps. Some theoretical results concerning this matter will be presented in Section \ref{sect:theoryASMC}.

To choose between different types of MCMC kernel is now a relatively straightforward extension of the above. Assume there are $I$ different MCMC kernels, each defined by a proposal distribution $q_{h,i}$, where $i \in \{1,\ldots,I\}$. The algorithm now learns a set of distributions, $\pi^{(t)}(h,i)$, for the pair of kernel type and associated tuning parameter. Each particle is assigned a random kernel type and tuning drawn form this distribution, with the pair, $(h_{t-1}^{(j)},i_{t-1}^{(j)})$,  associated with $\theta_{t-1}^{(j)}$. The algorithm proceeds by weighting this pair based on the observed $\tilde\Lambda(\theta_{t-1}^{(j)},\tilde\theta_t^{(j)})$ values as before, and updating the distribution, 
\begin{equation}\label{eqn:newkernelmixture}
    \pi^{(t)}(h,i) \propto \sum_{j=1}^M f(\tilde\Lambda(\theta_{t-1}^{(j)},\tilde\theta_t^{(j)})) R(h-h_{t-1}^{(j)})\delta_{i_{t-1}^{(j)}}(i).
\end{equation}
where $\delta_{i_{t-1}^{(j)}}(i)$ is a point mass on $i=i_{t-1}^{(j)}$. 
The method is described in detail below, see Algorithm \ref{alg:ASMC}. Within the specific implementation described, the sample of pairs, $(h,i)$, from $\pi^{(t)}(h,i)$ are allocated to particles randomly immediately after the resample--move step at iteration $t$. These pairs are then kept until the next iteration a resample--move step is called.

\begin{algorithm}
   \caption{The Adaptive SMC algorithm. Here, $\pi_0(\,\cdot\,),\ldots,\pi_n(\,\cdot\,)$ are an arbitrary sequence of targets; an MCMC kernel is assumed for particle dynamics.}
   \label{alg:ASMC}
   \begin{algorithmic}[1]
      \STATE Initialise from the prior $\{\theta_0^{(j)},w_0^{(j)}\}_{j=1}^M\sim\pi_0$.
      \STATE Draw a selection of pairs of MCMC kernels with associated tuning parameters, $\{(h_0^{(j)},K_{h,0}^{(j)})\}_{j=1}^M\equiv\{(h_0^{(j)},i_0^{(j)})\}_{j=1}^M\sim\pi(h,i)$, and attach one to each particle arbitrarily.
      \FOR{$t=1,\ldots,n$}
         \STATE Assume current $\{\theta_{t-1}^{(j)},w_{t-1}^{(j)}\}_{j=1}^M\sim\pi_{t-1}$
         \STATE Reweight $w_t^{(j)} = w_{t-1}^{(j)}\pi_t(\theta_{t-1}^{(j)})/\pi_{t-1}(\theta_{t-1}^{(j)})$. Result: $\{\theta_{t-1}^{(j)},w_t^{(j)}\}_{j=1}^M\sim\pi_t$.
         \IF{particle weights not degenerate (see text)}
            \STATE $\{\theta_{t}^{(j)},w_{t}^{(j)}\}_{j=1}^M \leftarrow \{\theta_{t-1}^{(j)},w_{t-1}^{(j)}\}_{j=1}^M$
            \STATE $\{(h_{t}^{(j)},K_{h,t}^{(j)})\}_{j=1}^M \leftarrow \{(h_{t-1}^{(j)},K_{h,t-1}^{(j)})\}_{j=1}^M$
            \STATE $t\rightarrow t+1$.
         \ELSE
            \STATE Resample: let $\mathcal{K}=\{k_1,\ldots,k_M\}\subseteq\{1,\ldots,M\}$ be the resampling indices, then $\{\theta_{t-1}^{(k)},1/M\}_{k\in\mathcal K}\sim\pi_t$. Relabel: $k_j \leftarrow j$, the $j$th resampling index so that $\{\theta_{t-1}^{(j)},1/M\}_{j=1}^M\sim\pi_t$. DO NOT resample kernels or tuning parameters at this stage.
            \STATE Move $\theta_{t-1}^{(j)}$ via the $\pi_t$--invariant MCMC kernel, $K_{h,t}^{(j)}$, and tuning parameter $h_{t-1}^{(j)}$, denote the proposed new particle as $\tilde\theta_t^{(j)}$ and accepted/rejected particle as $\theta_t^{(j)}$. Result: $\{\theta_t^{(j)},1/M\}_{j=1}^M\sim\pi_t$.
            \STATE To obtain $\{(h_{t}^{(j)},K_{h,t}^{(j)})\}_{k=1}^M\equiv\{(h_t^{(j)},i_t^{(j)})\}$, sample $M$ times from (\ref{eqn:newkernelmixture}). Allocate the new selection to particles at random.
         \ENDIF
      \ENDFOR
   \end{algorithmic}
\end{algorithm}

\section{Theoretical Results \label{sect:theoryASMC}}

In this section the proposed algorithm will be justified by a series of theoretical results; guidance as to how it should best be implemented will also be given. The results presented here apply in the limit as the number of particles, $M\rightarrow\infty$. As discussed above, in this limit, the variance of the kernel $R(\,\cdot\,)$ in (\ref{eq:pit}) tends to 0.

To simplify the discussion, it will be assumed that tunings are one dimensional (the arguments presented extend readily to the multivariate case). For a slight notational simplification, the criterion $\Lambda$ will be used, rather than $\tilde\Lambda$ (as suggested in algorithm \ref{alg:ASMC}); this does not affect the validity of any of the arguments, which also hold for $\tilde\Lambda$. The section is split into two parts.

Firstly, in section \ref{sect:onestep}, it is of interest to examine what happens to the distribution of the $h$s after one step of reweighting and resampling; this result will lead to a criterion for the choice of weight function that guarantees MCMC mixing improvement with respect to $\Lambda$. In section \ref{sect:manystep}, the sequential improvement of $h$s will be considered over many steps of the ASMC algorithm and with a changing target. General conditions for convergence of ASMC to the optimal kernel and tuning parameter will be provided.

\subsection{One Step Improvement and Weighting Function\label{sect:onestep}}

In this section and in the relevant proofs, it is appropriate to temporarily drop the $t$ superscript, eg $g^{(t)}\equiv g$, $\theta_{t-1}\equiv\theta$ and $\theta_t\equiv\theta'$. To study the effect of  reweighting and resampling on the distribution of the $h$s, suppose that currently $\{h^{(j)}\}_{j=1}^M\simiid\pi(h)$, the pdf of a random variable, $H$. The dependence on current and proposed particles means the weight attached to $h^{(j)}$ is random, but also, due to the independence of $h$ with the particles, is an unbiased estimator of the `true' weight, $\E_{\Theta,\Theta'|H}[f(\Lambda)|H=h^{(j)}]$, where $\E_{\Theta,\Theta'|H}$ denotes the expectation with respect to the joint density of the random variables $\Theta$ and $\Theta'$ conditional on $H$. The true \emph{weighting function} will be denoted,
\begin{equation}\label{eqn:weightfun}
   w(h) = \E_{\Theta,\Theta'|H}[f(\Lambda)|H=h].
\end{equation}
The following proposition, which is used repeatedly in subsequent results, shows how reweighting and resampling affects $\pi(h)$.

\begin{Propn}\label{propn:wboot}
   Suppose that currently $\{h^{(j)}\}_{j=1}^M\simiid\pi(h)$, the pdf of a random variable, $H$, independent of $\theta$. Let $w(h)$ be the weighting function defined as in (\ref{eqn:weightfun}). Then in the limit as $M\rightarrow\infty$, the distribution of the reweighted and subsequently resampled $h$s is,
   \[
      \pi^\star(h) = \frac{w(h)\pi(h)}{\int w(h)\pi(h)\rmd h}.
   \]
\end{Propn}

 \textbf{Proof:} See Appendix \ref{sect:proofwboot}.\qed

Since ASMC uses a selection of $h$s, it is appropriate as a starting point to look for conditions under which their \emph{distribution} is improved. It would be desirable if, over $\pi^\star(h)$, the objective function would on average take a higher value, for then the new distribution would on average perform better with respect to $\Lambda$ than the old. This criterion can be stated in mathematical form: conditions on $f$ are sought for which,
\begin{equation*}
   \int\pi^\star(h)g(h)\rmd h \geq \int\pi(h)g(h)\rmd h.
\end{equation*}

\begin{Lem}\label{lem:improve_criterion}
Assuming $g$ is $\pi(h)$--integrable, in the limit as $M\rightarrow\infty$,
\begin{equation}\label{eqn:cov_improve}
   \E_{\pi^\star(h)}[g(h)] \geq \E_{\pi(h)}[g(h)]\iff\cov_{\pi(h)}[g(h),w(h)] \geq 0.
\end{equation}
That is, provided there is positive correlation between the objective function $g(h)$ and the weighting function, $w(h)$, the new distribution of $h$s will on average perform better (on $g(h)$) with respect to $\Lambda$ than the old.
\end{Lem}
\textbf{Proof:} The result is obtained by expanding definitions in $(\ref{eqn:cov_improve})$:
\begin{eqnarray*}
   \E_{\pi^\star(h)}[g(h)] &\geq& \E_{\pi(h)}[g(h)],\\
   \iff\E_{\pi(h)}[w(h)g(h)] &\geq& \E_{\pi(h)}[w(h)]\E_{\pi(h)}[g(h)],\\
   \iff\cov_{\pi(h)}[g(h),w(h)] &\geq& 0.
\end{eqnarray*}\qed

Although this result does not directly yield a general form for $f$, it does give a simple criterion that must be fulfilled by any candidate function. An immediate corollary gives more concrete guidance:

\begin{Cor}
A simple linear weighting scheme, $f(\Lambda) = a + \Lambda$, where $a\geq0$, satisfies (\ref{eqn:cov_improve}). 
\end{Cor}
\textbf{Proof:} This is trivially verified using the linearity property of the covariance.\qed

A consequence of this lemma is that  the ASMC algorithm with linear weights will lead to sequential improvement with respect to $\Lambda$ under very weak assumptions on the target and initial density for $h$. 
A linear weighting scheme may at first glance seem sub--optimal, and that it should be possible to learn $h$ more quickly using a function $f(\Lambda)$ that increases at a super--linear rate. The present authors conjecture that such functions will not always guarantee an improvement in the distribution of $h$. For example consider $f(\Lambda) = \Lambda^2$, where the weighting function takes the form, $w(h) = g(h)^2 + \V[\Lambda|H=h]$. Because of the $\V[\Lambda|H=h]$ term, which may be large for values of $h$ where $g(h)$ is small, it is no longer true that $\cov_{\pi(h)}[g(h),w(h)]\geq0$ in general.

\subsection{Convergence Over a Number of Iterations\label{sect:manystep}}

The goal of this section is to provide a theoretical result concerning the ability of ASMC to update the distribution of $h$s with respect to a sequence of targets, $\pi_1(\theta_1),\ldots,\pi_n(\theta_n)$. To simplify notation, it will be assumed that a move occurs at each iteration of the algorithm. The result can be extended to the case where moves occur intermittently, providing they incur infinitely often in the limit as the number of data points goes to infinity. 

Define a set of functions, $\{g^{(t)}(h)\}_{t=1}^n$,
\begin{equation*}
    g^{(t)}(h) = \int \pi_t(\theta_{t-1})K_{h,t}(\theta_{t-1},\theta_t)\Lambda(\theta_{t-1},\theta_t)\rmd\theta_{t-1}\rmd\theta_t\geq0,
\end{equation*}
where, for each $t$, $K_{h,t}$ is a $\pi_t$--invariant MCMC kernel. 

For a linear weighting scheme,
\begin{equation*}
    \pi^{(t)}(h) \propto \pi(h)\prod\limits_{s=1}^t(a+g^{(s)}(h)).
\end{equation*}
Below it will be shown that as $t\rightarrow \infty$ if the sequence of functions, $\{g^{(t)}(h)\}$, converge to a fixed function, $g(h)$, and if $g$ has a unique global maximum, $\hopt$, then $\pi^{(t)}(h)$ will converge to a point mass on $\hopt$.

The key assumption of this theorem regards the convergence of the functions $\{g^{(t)}(h)\}$. This assumption is linked to the idea that a good value of $h$ for the target at time $t$ is required to be a good value at times later on. As mentioned above, the motivation behind SMC is that successive targets should be similar. Moreover, standard Bayesian asymptotic theory shows that as the number of observations, $n$, tends to infinity, the posterior tends in distribution to that of a Gaussian random variable. Thus, providing information from the current parameters about the posterior variance is used appropriately, it should be expected that the sequence of functions, $\{g^{(t)}(h)\}$, would also converge. This issue will be explored empirically in the next section.

\begin{Thm} \label{thm:convvari}
   Let $\pi(h)$ be the initial density for the tuning parameter with support $\mathcal{H}\subseteq\real$ and $a>0$. Define, as above, 
   \begin{equation*}
      \pi^{(t)}(h) \propto \pi(h)\prod\limits_{s=1}^t(a+g^{(s)}(h)).
   \end{equation*}
   Suppose there exists a function $g:\mathcal{H}\rightarrow\real_{\geq0}$ such that
   \begin{equation*}
      \sup_{h\in\mathcal H}|g^{(t)} - g| \leq k_gt^{-\alpha},\qquad \alpha\in(0,1),\ k_g>0.
   \end{equation*}
   Furthermore, suppose $g$ has a unique global maximum, $\hopt$, contained in the interior of $\mathcal H$ and that $g$ is twice differentiable in an interval containing $\hopt$. Then as $t\rightarrow\infty$, $\pi^{(t)}(h)$ tends to a Dirac mass centred on the optimal scaling, $\hopt$.
\end{Thm}

\textbf{Proof:} See Appendix \ref{sect:proofvari}.\qed

\section{Results \label{sect:resultsASMC}}

This section is organised as follows. In Section \ref{sect:convergeh}, the convergence of $h$ to an optimal scaling will be demonstrated empirically using a linear Gaussian model. Then in Section \ref{sect:bayesian_mixtures} the problem of Bayesian mixture analysis will be introduced. In Sections \ref{sect:detailsofSMCsims} and \ref{sect:resultsofsims} the proposed method will be evaluated in simulation studies using the example of Bayesian mixture posteriors as defining the sequence of targets of interest.

Following \cite{sherlock2009}, the expected (Mahalanobis) square jumping distance will be considered as an MCMC performance criterion:
\begin{equation*}
   \Lambda(\theta_{t-1},\theta_t) =  (\theta_{t-1}-\theta_t)^T\hat\Sigma_{\pi_t}\inv(\theta_{t-1}-\theta_t),
\end{equation*}
where $\theta_{t-1}$ and $\theta_t$ are two points in the parameter space and $\hat\Sigma_{\pi_t}$ is an empirical estimate of the target covariance obtained from the current set of particles. 

Two different MCMC kernels will be considered within the SMC algorithm, these are defined by the following two proposals:
\begin{eqnarray*}
   q_{\text{rw}}(\theta_{t-1},\tilde\theta_t) &=& \N(\theta_{t-1},h^2\hat\Sigma_{\pi_t}),\\
   q_{\text{lw}}(\theta_{t-1},\tilde\theta_t) &=& \N(\alpha\theta_{t-1} + (1-\alpha)\bar{\theta}_t,h^2\hat\Sigma_{\pi_t}), \qquad h\in(0,1],\ \alpha=\sqrt{1-h^2},
\end{eqnarray*}
where $\bar{\theta}_t$ and $\hat\Sigma_{\pi_t}$ are respectively estimates of the target and covariance. The first of these is a \emph{random--walk} proposal. The second is based upon a method for updating parameter values in \cite{SMCMPliuwest}, here named the `\emph{Liu/West}' proposal. 
The Liu/West proposal has mean shrunk towards the mean of the target and the imposed choice of $\alpha=\sqrt{1-h^2}$ sets the mean and variance of proposed particles to be the same as that of the current particles. Note that if the target is Gaussian, then this proposal can be shown to be equivalent to a Langevin proposal \citep{roberts1996}.

\subsection{Convergence of $h$\label{sect:convergeh}}

It is of interest to see an example $g(h)$ and demonstrate convergence of one of the proposed algorithms to the optimal scaling. This will be achieved using a Gaussian target, for which a useable analytic expression for the optimal scaling for the random walk kernel is available. The results in this section are based on 100 observations simulated from a 5--dimensional standard Gaussian density, $y_{1:100}\simiid\N(0,\I_5)$, where $\I_5$ is the $5\times5$ identity matrix. The observation variance was assumed to be known and therefore the probability model, or likelihood, was specified as,
\begin{equation*}
   \pi(y|\theta) = \N(y;\theta,\I_5).
\end{equation*}
The prior on the unknown parameter, $\theta$, the vector of means, was set to $\N(0,5\I_5)$. ASMC with a random walk proposal was used to generate $M=2000$ particles from the posterior. Resampling was invoked when the ESS dropped below $M/2$ and no noise was added to the $h$s after resampling. The initial distribution for $h$ was chosen to be uniform on $(0,10)$. Note that this model admits exact inference via the Kalman Filter.

\begin{figure}[htbp]
 \begin{minipage}{0.5\textwidth}
      \includegraphics[width=0.9\textwidth,height=0.9\textwidth]{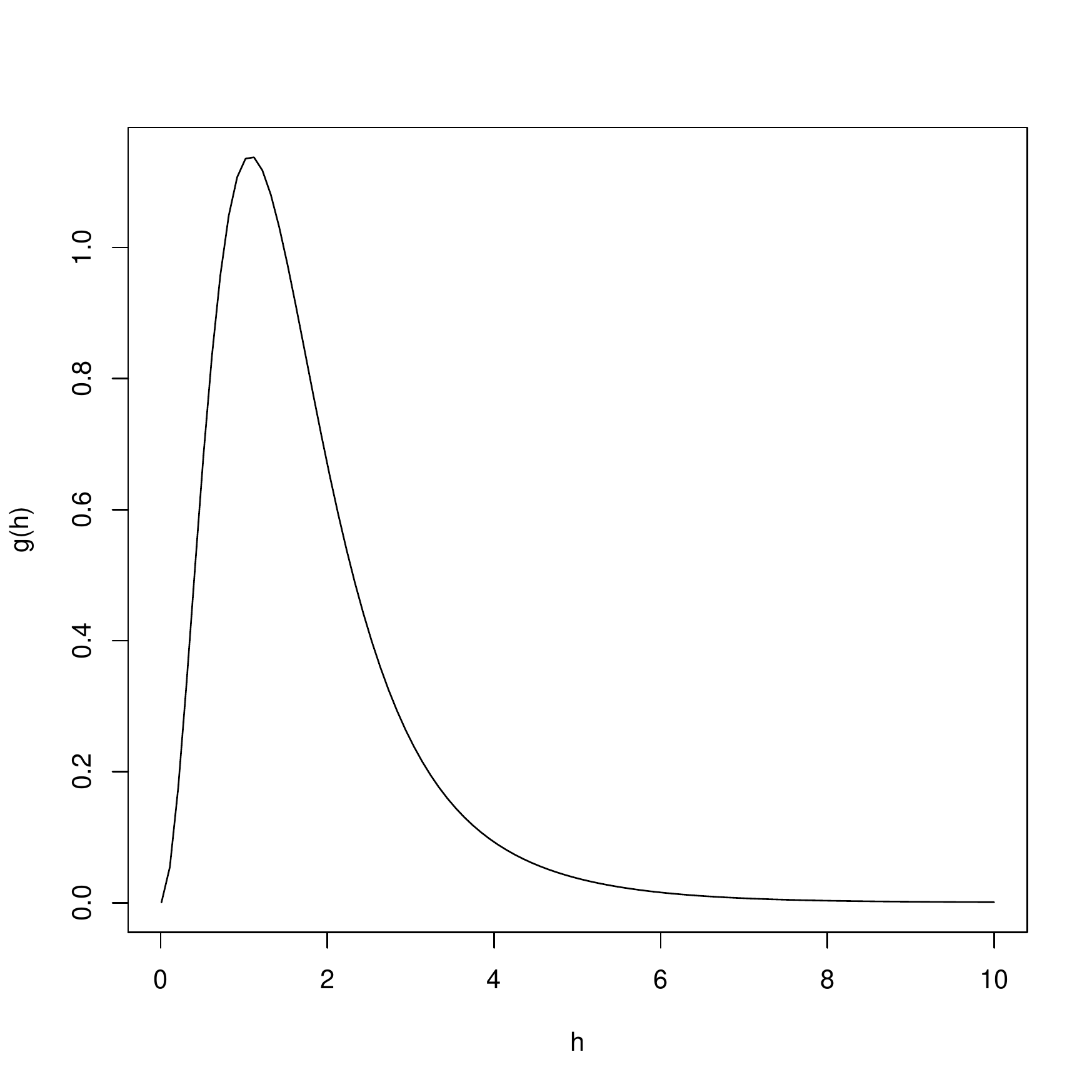}
   \end{minipage}
   \begin{minipage}{0.5\textwidth}
      \includegraphics[width=0.9\textwidth,height=0.9\textwidth]{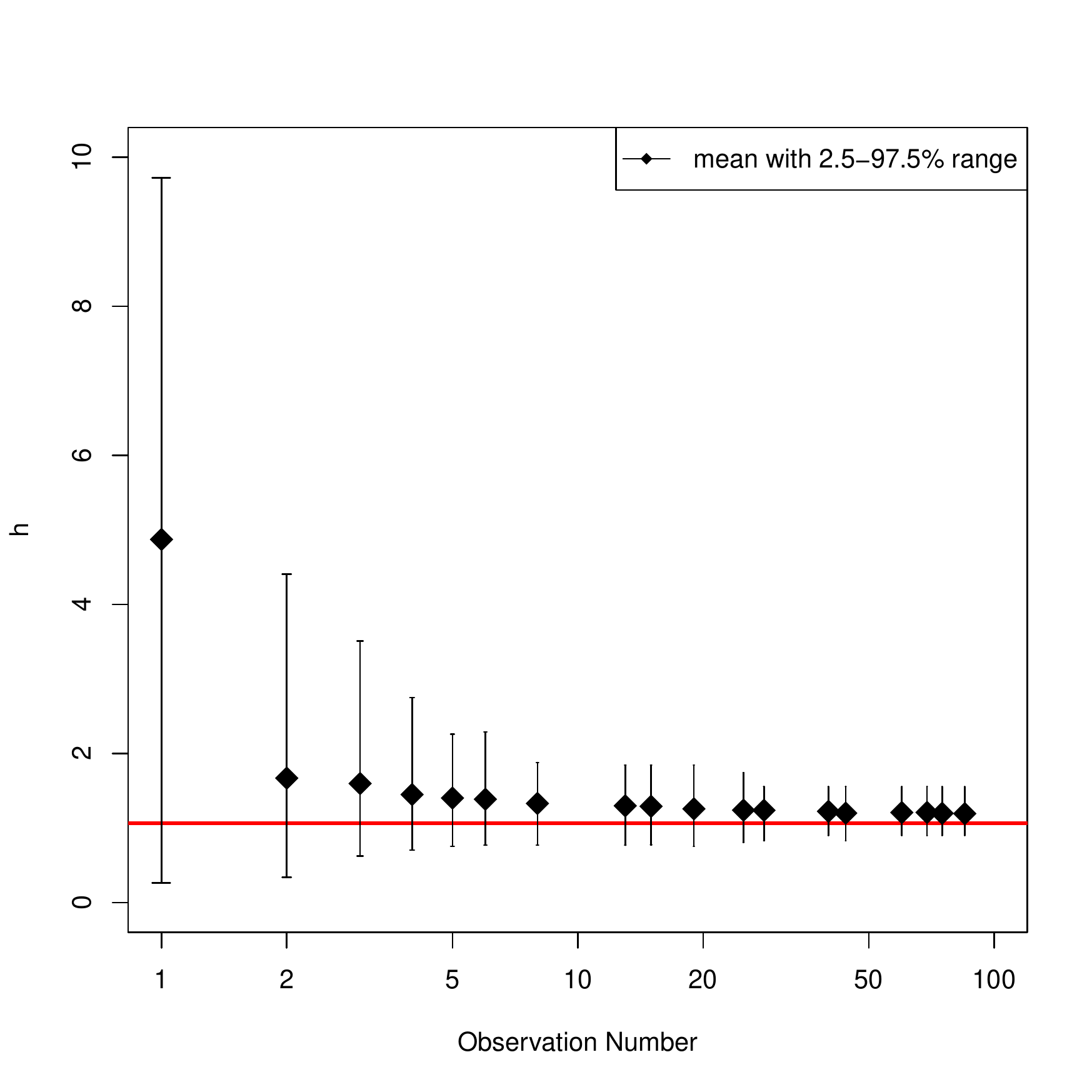}
   \end{minipage}
   \caption{\label{fig:gaussiansimresults}Left plot: $g(h)$ for a 5--dimensional Gaussian Target, explored with RWM and with ESJD as the optimization criterion. Right hand plot: convergence of $h$ for the same density based on 100 simulated observations; the horizontal line is the approximately optimal scaling, $1.06$.}
\end{figure}

The left hand plot in Figure \ref{fig:gaussiansimresults} shows $g(h)$ for this target. Note in this case that the sequence of functions, $\{g^{(t)}\}$, does not change much since each intermediate target is exactly Gaussian and the proposal is scaled by the approximate variance of the target. The optimum scaling, $\hopt$, was computed using 1-dimensional numerical integration and Theorem 1 of \cite{sherlock2009}. The right hand plot illustrates several features of the adaptive RWM; the resampling frequency, that the algorithm does indeed converge to the true optimal scaling and the approximate rate of this convergence.

\subsection{Bayesian Mixture Analysis \label{sect:bayesian_mixtures}}

The ability of the ASMC algorithm to learn MCMC tuning parameters in more complicated scenarios was evaluated using simulated data from mixture likelihoods (for a complete review of this topic, see \cite{fruhwirth-schnatter}). Let $p_1,\ldots,p_r>0$ be such that $\sum_{i=1}^rp_i=1$. Let $\N(\,\cdot\,;\mu,v)$ denote the normal density
function with mean $\mu$ and variance $v$. Let $\theta=\{p_{1:r-1},v_{1:r},\mu_{1:r}\}$.

The likelihood function for a single observation, $y_i$,is
\begin{equation}\label{eqn:mixlik}
   \pi(y_i|\theta) = \sum_{j=1}^rp_j\N(y_i;\mu_j,v_j).
\end{equation}
The prior $\theta$ was multivariate normal, on a transformed space using the generalised logit scale for the weights, log scale for variances, and leaving the means untransformed. The components of $\theta$ were assumed independent \emph{a priori}; the priors were $\log(p_j/p_r)\sim\N(0,1^2)$, $\log(v_j)\sim\N(-1.5,1.3^2)$ and $\mu_j\sim\N(0,0.75^2)$, where $j=1,\ldots,r-1$ in the case of the weights and $j=1,\ldots,r$ for the means and variances. The MCMC moves within the SMC algorithm were performed in the transformed space, using the appropriate inverse transformed values to compute the likelihood in (\ref{eqn:mixlik}).

An issue with mixture models is that for the above choice of prior, the likelihood and posterior are invariant to permutation of the component labels \citep{stephens2000}. As a result the posterior distribution has a multiple of $r!$ modes, corresponding to each possible permutation. One way of overcoming this problem is by introducing a constraint on the parameters, such as labelling the components so that $\mu_1<\mu_2<\cdots<\mu_r$, or so that $v_1<v_2<\cdots<v_r$. This choice will affect the empirical moments of the resulting posterior and hence the proposal distribution of the MCMC kernel -- both the random walk and Liu/West proposals depend on the posterior covariance, the latter also depending on the mean. In particular if there is a choice of ordering whereby the posterior is closer to Gaussian, then this is likely to lead to better mixing of the MCMC kernels. This phenomenon motivates the idea that it is also possible to choose between orderings on the parameter vector, which will be investigated in the sequel. 

\subsection{Details of Implementation of ASMC \label{sect:detailsofSMCsims}}

In analysing the simulated data, a number of SMC and ASMC algorithms were compared. These correspond to using the following MCMC kernels:
\begin{description}
   \item[RWfixed] Random walk ordered by means, with $h$ chosen based on the theoretical results for Gaussian targets \citep{roberts2001,sherlock2009}.
   \item[RWadaptive] Adaptive random walk ordered by means with uniform prior on $h$
   \item[LWmean] Adaptive Liu/West proposal ordered by means.
   \item[LWvariance] Adaptive Liu/West proposal ordered by variances.
   \item[Kmix] Adaptive choice between random walk ordered by means, Liu/West ordered on means and Liu/West ordered on variances.
\end{description}
In each case the reference to ordering relates to how the component labels were defined, and thus affect the estimate of the posterior mean and covariance used. 

The above methods were also compared with the adaptive MCMC algorithm of \cite{haario1998}, denoted \textbf{AMCMC}. The specific implementation used here is as follows. The prior densities were identical to those for ASMC, the parameter vector was ordered by means and the random walk tuning was computed using the approximately optimal Gaussian scaling of $h=2.4/\sqrt{3r-1}$. The AMCMC algorithm was run for 12000 iterations for the 5 dimensional datasets and for 30000 iterations for the 8 dimensional datasets: these values were chosen so as to approximately match the number of likelihood computations involved between the ASMC and AMCMC methods. The burn--in period was set to half of the number of iterations and the method was initialised by a draw from the prior. There was an initial non--adaptive phase, lasting 1000 iterations, where the proposal kernel was scaled by the prior covariance and after which scaling was via estimates of the posterior covariance computed from the chain to--date, this was updated every 100 iterations.

For the ASMC algorithms, the initial distribution of $h$s was chosen to be uniform on $(0,2)$ for the random walk and on $(0,1)$ for the Liu/West proposal. In the case of the random walk, this range of $h$s can be justified by considering the optimal scaling for a random walk Metropolis on a multivariate Gaussian target in $5$ dimensions namely $2.38/\sqrt{5}=1.06$ (and decays with increasing dimension as $O(d^{-1/2})$). For the Liu/West, $h$ must be in $(0,1]$. 

In each case a Gaussian kernel with variance $0.015^2$ was used in (\ref{eq:pit}). A sensitivity analysis showed the effect of changing the variance of the noise slightly did not affect the conclusions of this research. The parameter for the linear $h$-weighting scheme was $a=0$. If any $h$ was perturbed below zero, a small value, $1\times10^{-6}$, was imputed and similarly for the Liu/West approach, any $h$ perturbed above 1 was replaced by 1.

The number of particles was set to $M=2000$ for the 2--mixture datasets and $M=5000$ for the 3--mixture datasets.  Each algorithm was run 100 times on each dataset with the order of observations randomised each time. For the MCMC based methods an ESS tolerance of $M/2$ was used, as in \cite{jasra2007}. Resampling of the particles was via residual sampling \citep{whitley1994,liu1998}, but multinomial sampling was used in selecting $h$s. For ease of computing posterior quantities of interest, each of the above algorithms was forced to resample and move on the last iteration.

To compare the performance of different methods, a measure of the accuracy of the estimated predictive density was used. This is advantageous because it is invariant to re--labelling of the mixture components. The chosen accuracy measure was the variability of the predictive density (VPD) and was calculated as follows. Each run of the algorithm produces a weighted particle set, from which an estimate of $\E[\pi(y^{(i)}|y_{1:n})]$ can be obtained at 100 points, $\{y^{(i)}\}_{i=1}^{100}$, equi-spaced between -2.5 and 2.5. For each $i$, the 100 simulation runs produce 100 realisations of $\E[\pi(y^{(i)}|y_{1:n})]$; let $\hat y^{(i,j)}$ be the estimate of $y^{(i)}$ obtained from run $j$.  The VPD measure used in this paper is
\begin{equation*}
   \mathrm{mean}_i[\mathrm{var}_j(\hat y^{(i,j)})],
\end{equation*}
where $\mathrm{mean}_i$ is the mean over the $i$s and $\mathrm{var}_j$ is the variance of the estimates of $y^{(i)}$ obtained from the 100 simulations. The VPD gives an indication of the global variability of the predictive density across the simulations. In the tables, the relative VPD is used, which gives a scale--free comparison between methods. The SMC/ASMC algorithm with a relative VPD of 1 is the reference algorithm and has the smallest VPD of the SMC/ASMC methods; larger values indicate higher VPDs. For the AMCMC methods, the predictive densities were computed using all available samples ie with 6000 for the 2--mixture datasets and 15000 for the 3--mixture datasets. For the SMC/ASMC methods a Rao--Blackwellised version of the predictive density was computed using all current and proposed particles available from the last iteration (that is, using 4000/10000 sample points respectively for the 2/3--mixture datasets).

\subsection{Results \label{sect:resultsofsims}}

100 realisations from were simulated from the following likelihoods:
\begin{eqnarray*}
   \text{Dataset 1:} & & \pi(y|\mb\theta) = 0.5\N(y;-0.25,0.5^2) + 0.5\N(y;0.25,0.5^2),\\
   \text{Dataset 2:} & & \pi(y|\mb\theta) = 0.5\N(y;0,1^2) + 0.5\N(y;0,0.1^2),\\
   \text{Dataset 3:} & & \pi(y|\mb\theta) = 0.3\N(y;-1,0.5^2) + 0.7\N(y;1,0.5^2),\\
   \text{Dataset 4:} & & \pi(y|\mb\theta) = 0.5\N(y;-0.75,0.1^2) + 0.5\N(y;0.75,0.1^2),\\
   \text{Dataset 5:} & & \pi(y|\mb\theta) = \T0.35\N(y;-0.1,0.1^2) + 0.3\N(y;0,0.5^2) + 0.35\N(y;0.1,1^2),\\
   \text{Dataset 6:} & & \pi(y|\mb\theta) = \T0.25\N(y;-0.5,0.1^2) + 0.5\N(y;0,0.2^2) + 0.25\N(y;0.5,0.1^2),\\
\end{eqnarray*}

This choice of datasets in combination with the selection of MCMC kernels allows several hypotheses to be tested empirically. Firstly, by comparing the performance of RWfixed with RWadaptive in these cases, it is possible to see whether anything is lost or gained by adapting the proposal kernel. Secondly, the impact of the different kernel orderings on MCMC mixing will become apparent by considering the performance of LWmean and LWvariance in these settings. Datasets 1, 3, 4 and 6 have well `separated' means and similar variances, so one might expect algorithms ordering by means to perform better; whereas datasets 2 and 5 have well separated variances and similar means, so perhaps the algorithms ordering by variances might do well here. Thirdly, the Kmix algorithm should be able to choose the best ordering and it is of interest to compare the results from this algorithm with an adaptive version of the individual kernels. 

The simulation results from these datasets are presented in Table \ref{tab:adaptive_1-4}. These give both the relative VPD for each method, but also an estimated mean ESJD for each method.

\begin{table}
    \centering

   \footnotesize

\vspace{0.5em}
\textbf{Dataset 1}
\vspace{0.5em}

    \begin{tabular}{p{2.5cm}p{1.5cm}p{2cm}p{2cm}p{1.8cm}p{1.8cm}}
         Method & Rel. VPD & JD & Acc. & $h$ & Propn. \\ \hline
        LWvariance & 1 & 1.869 & 0.3 & 0.941 &  \\
        LWmean & 1.189 & 1.818 & 0.32 & 0.956 &  \\
        Kmix & 1.258 & 1.845 & 0.317 & LWm 0.963 LWv 0.958 & LWm 0.785 LWv 0.215 \\
        RWadaptive & 2.391 & 0.708 & 0.21 & 0.946 &  \\
        AMCMC & 2.396 & 0.575 & 0.13 & 1.073 & $\cdot$ \\
        RWfixed & 3.414 & 0.641 & 0.18 & 1.064 &  \\ \hline
    \end{tabular}

\vspace{0.5em}
\textbf{Dataset 2}
\vspace{0.5em}

    \begin{tabular}{p{2.5cm}p{1.5cm}p{2cm}p{2cm}p{1.8cm}p{1.8cm}}
        LWvariance & 1 & 9.139 & 0.873 & 0.978 &  \\
        Kmix & 2.843 & 9.023 & 0.854 & LWm 0.984 LWv 0.978 & LWm 0.005 LWv 0.995 \\
        AMCMC & 28.333 & 0.197 & 0.019 & 1.073 & $\cdot$ \\
        LWmean & 112.23 & 1.869 & 0.129 & 0.969 &  \\
        RWadaptive & 188.094 & 0.77 & 0.134 & 0.584 &  \\
        RWfixed & 219.907 & 0.596 & 0.041 & 1.064 &  \\ \hline
    \end{tabular}

\vspace{0.5em}
\textbf{Dataset 3}
\vspace{0.5em}

    \begin{tabular}{p{2.5cm}p{1.5cm}p{2cm}p{2cm}p{1.8cm}p{1.8cm}}
        LWmean & 1 & 6.38 & 0.792 & 0.98 &  \\
        Kmix & 1.54 & 6.378 & 0.806 & LWm 0.979 & LWm 1 \\
        AMCMC & 7.465 & 0.847 & 0.146 & 1.073 & $\cdot$ \\
        RWfixed & 40.538 & 1.124 & 0.277 & 1.064 &  \\
        RWadaptive & 45.739 & 1.057 & 0.369 & 1.045 &  \\
        LWvariance & 148.827 & 0.737 & 0.064 & 0.966 &  \\ \hline
    \end{tabular}

\vspace{0.5em}
\textbf{Dataset 4}
\vspace{0.5em}

    \begin{tabular}{p{2.5cm}p{1.5cm}p{2cm}p{2cm}p{1.8cm}p{1.8cm}}
         LWmean & 1 & 7.132 & 0.875 & 0.98 &  \\
        Kmix & 1.099 & 7.127 & 0.877 & LWm 0.979 & LWm 1 \\
        AMCMC & 24.024 & 0.462 & 0.057 & 1.073 & $\cdot$ \\
        RWadaptive & 48.606 & 1.143 & 0.274 & 1.086 &  \\
        RWfixed & 51.919 & 1.167 & 0.298 & 1.064 &  \\
        LWvariance & 1096.167 & 0.632 & 0.027 & 0.961 &  \\ \hline
    \end{tabular}

\vspace{0.5em}
\textbf{Dataset 5}
\vspace{0.5em}

    \begin{tabular}{p{2.5cm}p{1.5cm}p{2cm}p{2cm}p{1.8cm}p{1.8cm}}
        AMCMC & 0.883 & 0.356 & 0.04 & 0.849 & $\cdot$ \\
        Kmix & 1 & 2.258 & 0.234 & LWm 0.964 LWv 0.971 & LWm 0.044 LWv 0.956 \\
        LWvariance & 1.151 & 2.284 & 0.183 & 0.971 &  \\
        LWmean & 2.792 & 1.007 & 0.092 & 0.961 &  \\
        RWadaptive & 4.923 & 0.847 & 0.205 & 0.435 &  \\
        RWfixed & 5.187 & 0.56 & 0.055 & 0.84 &  \\ \hline
    \end{tabular}

\vspace{0.5em}
\textbf{Dataset 6}
\vspace{0.5em}

    \begin{tabular}{p{2.5cm}p{1.5cm}p{2cm}p{2cm}p{1.8cm}p{1.8cm}}
        LWmean & 1 & 4.099 & 0.277 & 0.972 &  \\
        Kmix & 1.018 & 3.994 & 0.363 & LWm 0.973 & LWm 1 \\
        AMCMC & 1.556 & 0.211 & 0.04 & 0.849 & $\cdot$ \\
        RWfixed & 3.244 & 0.996 & 0.429 & 0.84 &  \\
        RWadaptive & 3.259 & 0.93 & 0.192 & 0.693 &  \\
        LWvariance & 3.951 & 1.951 & 0.13 & 0.944 &  \\ \hline
    \end{tabular}
   
\caption{\label{tab:adaptive_1-4} Rel. VPD is relative VPD, JD is the mean square jumping distance, Acc is the mean final acceptance probability, $h$ is the mean final scaling by kernel and Propn is the mean final kernel proportions. The kernels `LWm' and `LWv' indicate respectively a Liu/West proposal ordering on means or variances.}

\end{table}

As would be hoped, a very strong correlation between lower VPD and higher ESJD is evident for the SMC/ASMC algorithms, this empirically supports the use of ESJD as the chosen criteria for adapting the MCMC kernels.

There is relatively little difference across scenarios between the fixed and adaptive random walk methods. Furthermore, the adaptive random walk settles on a similar scaling as the fixed scaled version in datasets 3 and 4, whereas in datasets 1, 2, 5 and 6, RWadaptive settles to values below RWfixed. In datasets 1, 2, 4 and 5, the adaptive RW outperformed the fixed equivalent (though the difference was negligible in datasets 4 and 5); this is likely due to the fact that the covariance was not a good estimate and the adaptive version of the algorithm was able to rescale to compensate for this. In datasets 3 and 6, the fixed random walk marginally outperformed the adaptive.

The `correctly ordered' sequential Liu/West algorithms considerably outperform the sequential RW--based methods in all six datasets and the incorrectly ordered versions perform worse or as poorly as the RW. For the Liu/West proposals, the $h$ selected in each dataset was very close to 1: this special value corresponds to an independence kernel in the form of a moment--matched Gaussian approximation of the target. This is of interest as, in combination with the high acceptance rates of between $80$--$87$\% in datasets 2--4, suggests that the `correct' ordering makes the target, ostensibly a very \emph{complex} density function, approximately Gaussian in these cases.

The Kmix algorithm is able to choose between orderings; the advantages of this are clearly evidenced in the results, as it selects the best ordering in each case, with the exception of dataset 1 (where the means and variances are both similar). The Kmix sampler settles almost unanimously on one ordering above the others. These results show empirically that there is not much difference in using a single (correctly chosen) kernel compared with using a selection of kernels.

The performance of AMCMC was surpassed in all cases by the Kmix algorithm with the exception of dataset 5, where AMCMC was the best performing algorithm. In this latter case and in dataset 6, neither AMCMC nor the SMC/ASMC algorithms performed well. AMCMC outperformed RWadaptive in each case apart from in dataset 1, where the difference was small. However, the results show the average jumping distance of the kernel used in the ASMC algorithm was greater than that of AMCMC in all cases, suggesting ASMC is able to adapt better to well-mixing kernels. To make this comparison more clear, two MCMC algorithms were run on each data-set, one using the final kernel found by AMCMC and one using a kernel based on the ASMC run, with the final estimated covariance matrix and the final mean value of the tuning parameter. The resulting MCMC algorithms performed very similarly in 3 cases (VPD of the two MCMC algorithms within 10\% of each other) and the kernel found by ASMC performed better in the other 3 (VPD reduced by 30\%, 40\% and 80\%).

\section{Discussion \label{sect:discuss_extASMC}}

This paper introduces a new method for automatically tuning and choosing between different MCMC kernels. Where MCMC based SMC code already exists, adapting the $h$s would be a relatively straightforward means of enhancing performance, the main effort being in calculating the ratio of the proposed particles in the accept/reject step. Probably the most important conclusion from the simulation studies presented is that there is not much lost in terms of performance in the adaption process -- the Kmix algorithm performed comparably to the respective best performing individual component and the adaptive random walk Metropolis performed similarly to the fixed, approximately optimally scaled version.

Although the method as presented has assumed that i.i.d. observations are available from the likelihood, ASMC readily extends to the case of a dependent sequence. Furthermore, the extension to general sequences of target densities is immediate, and implied by the choice of notation in Algorithm \ref{alg:ASMC}. The theoretical results presented in section \ref{sect:theoryASMC} only apply to a one--dimensional $h$, in the case that the tuning parameter is a vector, the proposed algorithm and theoretical results still apply (with slight modifications), but convergence is likely to be at a slower rate.

The main assumption of ASMC is that a good $h$ at time $t$ is likely also to perform well at time $t+1$. One piece of evidence that supports this assumption is that the resampling frequency decreases with an increasing number of observations \citep{chopin2002}. This implies that, although $\pi_1$ and $\pi_2$ may be quite different, $\pi_{1001}$ and $\pi_{1002}$ are likely to be less so, provided that the data provides sufficient information on the parameters. As mentioned earlier in the text, the assumption of similar successive target densities is also required for the non--adaptive version \citep{chopin2002,delmoral2006}.

ASMC can be easily extended by considering other proposal densities. For example it is possible to formulate a $T$--distributed version of the Liu/West proposal, this allows for heavier tailed proposals, the heaviness of which can be selected automatically by adaptively choosing the number of degrees of freedom; this $t$--based proposal includes the Liu/West as a special case. Other interesting algorithms can be formulated using DE proposals \citep{terbraak2006} (which generalises the snooker algorithm of \cite{gilks1994}) or regional MCMC proposals \citep{roberts2009,craiu2009} -- both of which appeal strongly to the particle structure of the new method.

\appendix

\section{Proof of Proposition \ref{propn:wboot} \label{sect:proofwboot}}

Let $\Lambda^{(j)} = \Lambda(\theta^{(j)},\theta'^{(j)})$ ie the observed $\Lambda$ for the $j$th particle and $\I$ denote the indicator function. The collection $\{h^{(j)},1/M\}_{j=1}^M$ is an iid sample from $\pi(h)$, but with weights defined as,
\begin{equation*}
   W^{(j)} = \frac{f(\Lambda^{(j)})}{\sum_{i=1}^M f(\Lambda^{(i)})},
\end{equation*}
the weighted particle set, $\{h^{(j)},W^{(j)}\}_{j=1}^M$, has empirical density,
\begin{equation*}
   \tilde\pi(h) = \sum_{j=1}^M W^{(j)}\I(h=h^{(j)}).
\end{equation*}
Define a discrete random variable $H^\star$, which takes value $h^{(j)}$ with probability $W^{(j)}$. For $h^\star\in\real$,
\begin{equation*}
   \prob(H^\star \leq h^\star) = \sum_{j=1}^M W^{(j)}\I(h^{(j)}\leq h^\star) = \frac{\frac1M\sum_{j=1}^M f(\Lambda^{(j)})\I(h^{(j)}\leq h^\star)}{\frac1M\sum_{i=1}^M f(\Lambda^{(i)})}.
\end{equation*}
In the limit as $M\rightarrow\infty$, $(\theta,\theta')\simiid\pi(\theta)K(\theta,\theta')$ the strong law of large numbers implies,
\begin{eqnarray*}
   \frac{\frac1M\sum_{j=1}^M f(\Lambda^{(j)})\I(h^{(j)}\leq h^\star)}{\frac1M\sum_{i=1}^M f(\Lambda^{(i)})} &\rightarrow& \frac{\E_{\pi(\theta)K(\theta,\theta')}[f(\Lambda)\I(H<h^\star)]}{\E_{\pi(\theta)K(\theta,\theta')}[f(\Lambda)]}, \\
   &=& \frac{\E_{H}\left[\E_{\Theta,\Theta'|H}[f(\Lambda)|H=h]\I(H<h^\star)\right]}{\E_{H}\left[\E_{\Theta,\Theta'|H}[f(\Lambda)|H=h]\right]}
\end{eqnarray*}
using the properties of conditional expectation. To complete the proof, observe that $\E_{\Theta,\Theta'|H}[f(\Lambda)|H=h]=w(H)$, so,
\begin{equation*}
   \lim_{M\rightarrow\infty}\prob(H^\star\leq h^\star) = \frac{\E_{\pi(h)}\left[w(H)\I(H<h^\star)\right]}{\E_{\pi(h)}\left[w(H)\right]} =\frac{\int_{s\leq h^\star}w(s)\pi(s)\rmd s}{\int w(h)\pi(h)\rmd h};
\end{equation*}
convergence in distribution follows as required. \qed

\section{Proof of Theorem \ref{thm:convvari} \label{sect:proofvari}}

The proof proceeds in two parts and starts by observing that $\pi^{(n)}(h) = \pi(h)\exp\{nf_n\}$ where,
\begin{equation*}
   f_n(h) = \frac1n\sum_{t=1}^n\log(a+g^{(t)}(h)),
\end{equation*}
In the first part, the following results will be proved:
\begin{itemize}
   \item \label{convvari:part1} There exists a function, $f:\mathcal{H}\rightarrow\real$, such that $\sup_{h\in\mathcal{H}}|f_n - f| < k_fn^{-\alpha}$.
   \item \label{convvari:part2} $\hopt$ is the unique global maximum of $f$.
   \item $f$ is twice differentiable in an interval containing $\hopt$.
\end{itemize}
In the second part of the proof, these results will be used to show that as $n\rightarrow\infty$, $\pi^{(n)}(h)$ approaches a Dirac mass centred on $\hopt$.

\textbf{Part 1}

Claim that $f(h) = \log(a+g(h))$. It is easy to show $\sup_{h\in\mathcal{H}}|(a+g^{(t)})/(a+g) - 1|<k_lt^{-\alpha}$,
where $k_l=k_g/\inf_{h\in\mathcal{H}}\{a+g(h)\}<\infty$ by assumption. Put $k_m=k_l+1> k_l+k_lt^{-\alpha}/2$ for all $t>\exp\{-2/k_l\alpha\}$. For sufficiently large (finite) $t$ and any $h\in\mathcal{H}$, $(a+g^{(t)})/(a+g)$ is close to 1, a Taylor series argument can therefore be applied to give,
\begin{equation*}
   -k_mt^{-\alpha} < -k_lt^{-\alpha} - k_lt^{-2\alpha}/2 < \log(1-k_lt^{-\alpha}) < \log[(a+g^{(t)})/(a+g)] < k_lt^{-\alpha}.
\end{equation*}
The preceding argument shows that,
\begin{equation*}
   \sup_{h\in\mathcal{H}}\left|\log\left\{\frac{a+g^{(t)}}{a+g}\right\}\right|=\sup_{h\in\mathcal{H}}|\log(a+g^{(t)}) - \log(a+g)|< k_mt^{-\alpha}.
\end{equation*}
For all $h\in\mathcal{H}$,
\begin{eqnarray*}
   \left|f_n - \log(a+g)\right| &<& \frac1n\sum_{t=1}^n\left|\log(a+g^{(t)}) - \log(a+g)\right|,\\
   &<& \frac{k_m}n\sum_{t=1}^nt^{-\alpha},\\
   &<& k_fn^{-\alpha},
\end{eqnarray*}
as required. If $g$ is twice differentiable in an interval $\mathcal{I}\subseteq\mathcal{H}$, where $\mathcal{I}\ni\hopt$ then, being a continuous function of $g$, $f$ is also twice differentiable on $\mathcal{I}$. That $\hopt$ is the unique global maximum of $f$ is now implied by the assumptions on $g$ and the strict increasing monotonicity of the logarithm.

\textbf{Part 2}

In this part, the properties of $f$ will be used to show that for any interval containing $\hopt$ as an interior point and as $n\rightarrow\infty$, the probability that $h$ belongs to that interval tends to 1.

Let $\bar{\mathcal X}$ denote the compliment of $\mathcal X$ in $\mathcal H$. Let $\mathcal I_0\supset\mathcal H$ be any interval containing $\hopt$ as an interior point. By virtue of the global uniqueness of $\hopt$, there exists an open interval $\mathcal I_1\supset\mathcal I_0$ also containing $\hopt$ such that $f''(h)<0$ for all $h\in\mathcal I_1$ and with the property, $\inf_{h\in\mathcal I_1}f\geq\sup_{h\in\bar{\mathcal I}_1}f$.

Then for all open intervals $\mathcal I_2\supset\mathcal I_1$, define,
\begin{eqnarray*}
   \sup_{h\in\mathcal I_2}\{f(\hopt)-f(h)\}=\epsilon_1,\\
   \inf_{h\in\bar{\mathcal I}_1}\{f(\hopt)-f(h)\}=\epsilon_2.
\end{eqnarray*}
The strict concavity of $f$ on $\mathcal I_1$ implies $\epsilon_2>\epsilon_1$ (note the strict inequality). Consider the probability of $h\in\mathcal I_0$ after $n$ updates,
\begin{eqnarray*}
    \prob(h\in\mathcal I_0) > \prob(h\in\mathcal I_1) = \frac{\int_{\mathcal I_1}\pi^{(n)}(h)\rmd h}{\int_{\mathcal H}\pi^{(n)}(h)\rmd h} &=& \frac{\int_{\mathcal I_1}\pi(h)\exp\{nf_n(h)\}\rmd h}{\int_{\mathcal H}\pi(h)\exp\{nf_n(h)\}\rmd h},\\
    &>& \frac{1}{ 1 + \frac{\int_{\bar{\mathcal I}_1}\pi(h)\exp\{nf_n(h)\}\rmd h}{\int_{\mathcal I_2}\pi(h)\exp\{nf_n(h)\}\rmd h}},
\end{eqnarray*}
since for any positive reals $a_1$, $a_2$ and $a_3$, if $a_1>a_2$ then $\frac{a_1}{a_1+a_3}>\frac{a_2}{a_2+a_3}\equiv\frac{1}{1+a_3/a_2}$. By uniform convergence of $f_n$, the quotient of integrals in the denominator can be bounded above by,
\begin{eqnarray*}
   \frac{\int_{\bar{\mathcal I}_1}\pi(h)\exp\{nf_n(h)\}\rmd h}{\int_{\mathcal I_2}\pi(h)\exp\{nf_n(h)\}\rmd h} &\leq& \frac{\int_{\bar{\mathcal I}_1}\pi(h)\exp\{nf(h) + k_fn^{1-\alpha}\}\rmd h}{\int_{\mathcal I_2}\pi(h)\exp\{nf(h) - k_fn^{1-\alpha}\}\rmd h},\\
     &\leq& \frac{\int_{\bar{\mathcal I}_1}\pi(h)\exp\{nf(\hopt) - n\epsilon_2 + k_fn^{1-\alpha}\}\rmd h}{\int_{\mathcal I_2}\pi(h)\exp\{nf(\hopt) - n\epsilon_1 - k_fn^{1-\alpha}\}\rmd h}, \\
    &\leq& \frac{\prob_{\pi(h)}(h\in\bar{\mathcal I}_1)}{\prob_{\pi(h)}(h\in{\mathcal I_2})}\exp\{n(\epsilon_1-\epsilon_2)+2k_fn^{1-\alpha}\},\\
   &\rightarrow& 0 \text{ as } n\rightarrow\infty,
\end{eqnarray*}
since $\epsilon_1-\epsilon_2 < 0$. Therefore $\prob(h\in\mathcal I_0) \rightarrow 1$ as $n\rightarrow\infty$. Since the choice of $\mathcal I_0\ni\hopt$ was arbitrary, it may be made infinitesimally narrow and still, after enough iterations of the sampler $\prob(h\in\mathcal I_0) \rightarrow 1$. This implies that $\pi^{(n)}(h)$ tends in distribution to a Dirac mass centred on $\hopt$ and establishes the claim. \qed

  \bibliographystyle{Chicago}
  \bibliography{bibliography.bib}

\begin{thebibliography}{}

\bibitem[\protect\citeauthoryear{Andrieu and Robert}{Andrieu and
  Robert}{2001}]{andrieu2001}
Andrieu, C. and C.~Robert (2001).
\newblock Controlled {MCMC} for optimal sampling.
\newblock Technical report, Universit\'e Paris--Dauphine.

\bibitem[\protect\citeauthoryear{Andrieu and Thoms}{Andrieu and
  Thoms}{2008}]{andrieu2008}
Andrieu, C. and J.~Thoms (2008).
\newblock A tutorial on adaptive {MCMC}.
\newblock {\em Statistics and Computing\/}~{\em 18\/}(4), 343--373.

\bibitem[\protect\citeauthoryear{Atchad\'e and Rosenthal}{Atchad\'e and
  Rosenthal}{2005}]{atchade2005}
Atchad\'e, Y. and J.~Rosenthal (2005).
\newblock On adaptive {M}arkov chain {M}onte {C}arlo algorithms.
\newblock {\em Bernoulli\/}~{\em 11\/}(5), 815--828.

\bibitem[\protect\citeauthoryear{Capp\'{e}, Douc, Guillin, Marin, and
  Robert}{Capp\'{e} et~al.}{2008}]{cappe2008}
Capp\'{e}, O., R.~Douc, A.~Guillin, J.-M. Marin, and C.~P. Robert (2008).
\newblock Adaptive importance sampling in general mixture classes.
\newblock {\em Statistics and Computing\/}~{\em 18\/}(4), 447--459.

\bibitem[\protect\citeauthoryear{Chopin}{Chopin}{2002}]{chopin2002}
Chopin, N. (2002).
\newblock A sequential particle filter method for static models.
\newblock {\em Biometrika\/}~{\em 89\/}(3), 539--552.

\bibitem[\protect\citeauthoryear{Cornebise, Moulines, and Olsson}{Cornebise
  et~al.}{2008}]{cornebise2008}
Cornebise, J., E.~Moulines, and J.~Olsson (2008).
\newblock Adaptive methods for sequential importance sampling with application
  to state space models.
\newblock {\em Statistics and Computing\/}~{\em 18\/}(4), 461--480.

\bibitem[\protect\citeauthoryear{Craiu, Rosenthal, and Yang}{Craiu
  et~al.}{2009}]{craiu2009}
Craiu, R.~V., J.~Rosenthal, and C.~Yang (2009).
\newblock Learn from thy neighbor: Parallel-chain and regional adaptive {MCMC}.
\newblock {\em Journal of the American Statistical Association\/}~{\em
  104\/}(488), 1454--1466.

\bibitem[\protect\citeauthoryear{Del~Moral, Doucet, and Jasra}{Del~Moral
  et~al.}{2006}]{delmoral2006}
Del~Moral, P., A.~Doucet, and A.~Jasra (2006).
\newblock Sequential {M}onte {C}arlo samplers.
\newblock {\em Journal of the Royal Statistical Society: Series B (Statistical
  Methodology)\/}~{\em 68\/}(3), 411--436.

\bibitem[\protect\citeauthoryear{Douc, Guillin, Marin, and Robert}{Douc
  et~al.}{2005}]{douc2005a}
Douc, R., A.~Guillin, J.-M. Marin, and C.~P. Robert (2005).
\newblock Minimum variance importance sampling via population {M}onte {C}arlo.
\newblock Technical report.

\bibitem[\protect\citeauthoryear{Doucet, de~Freitas, and Gordon}{Doucet
  et~al.}{2001}]{SMCMiP}
Doucet, A., N.~de~Freitas, and N.~Gordon (Eds.) (2001).
\newblock {\em Sequential {M}onte {C}arlo Methods in Practice}.
\newblock Springer--Verlag New York.

\bibitem[\protect\citeauthoryear{Fearnhead}{Fearnhead}{2002}]{fearnhead2002}
Fearnhead, P. (2002).
\newblock {MCMC}, sufficient statistics and particle filters.
\newblock {\em Journal of Computational and Graphical Statistics\/}~{\em 11},
  848--862.

\bibitem[\protect\citeauthoryear{Fearnhead}{Fearnhead}{2008}]{fearnhead2008}
Fearnhead, P. (2008).
\newblock Computational methods for complex stochastic systems: A review of
  some alternatives to {MCMC}.
\newblock {\em Statistics and Computing\/}~{\em 18}, 151--171.

\bibitem[\protect\citeauthoryear{Fr\"uhwirth-Schnatter}{Fr\"uhwirth-Schnatter}%
{2006}]{fruhwirth-schnatter}
Fr\"uhwirth-Schnatter, S. (2006).
\newblock {\em Finite Mixture and {M}arkov Switching Models}.
\newblock Springer.

\bibitem[\protect\citeauthoryear{Gamerman and Lopes}{Gamerman and
  Lopes}{2006}]{gamermanlopes}
Gamerman, D. and H.~F. Lopes (2006).
\newblock {M}arkov chain {M}onte {C}arlo: Stochastic simulation for {B}ayesian
  inference (2nd ed.).

\bibitem[\protect\citeauthoryear{Gilks and Berzuini}{Gilks and
  Berzuini}{1999}]{gilks2001}
Gilks, W. and C.~Berzuini (1999).
\newblock Following a moving target -- {M}onte {C}arlo inference for dynamic
  {B}ayesian models.
\newblock {\em {Journal of the Royal Statistical Society, Series B}\/}~{\em
  63\/}(1), 127--146.

\bibitem[\protect\citeauthoryear{Gilks, Richardson, and Spiegelhalter}{Gilks
  et~al.}{1995}]{MCMCiP}
Gilks, W., S.~Richardson, and D.~Spiegelhalter (Eds.) (1995).
\newblock {\em {M}arkov Chain {M}onte {C}arlo in Practice}.
\newblock Chapman \& Hall/CRC.

\bibitem[\protect\citeauthoryear{Gilks, Roberts, and George}{Gilks
  et~al.}{1994}]{gilks1994}
Gilks, W.~R., G.~O. Roberts, and E.~I. George (1994).
\newblock Adaptive direction sampling.
\newblock {\em Journal of the Royal Statistical Society. Series D (The
  Statistician)\/}~{\em 43\/}(1), 179--189.

\bibitem[\protect\citeauthoryear{Gordon, Salmond, and Smith}{Gordon
  et~al.}{1993}]{gordon1993}
Gordon, N.~J., D.~J. Salmond, and A.~F.~M. Smith (1993).
\newblock Novel approach to nonlinear/non-{G}aussian {B}ayesian state
  estimation.
\newblock {\em Radar and Signal Processing, IEE Proceedings F\/}~{\em
  140\/}(2), 107--113.

\bibitem[\protect\citeauthoryear{Haario, Saksman, and Tamminen}{Haario
  et~al.}{1998}]{haario1998}
Haario, H., E.~Saksman, and J.~Tamminen (1998).
\newblock An adaptive {M}etropolis algorithm.
\newblock {\em Bernoulli\/}~{\em 7}, 223--242.

\bibitem[\protect\citeauthoryear{Hastings}{Hastings}{1970}]{hastings1970}
Hastings, W.~K. (1970).
\newblock {M}onte {C}arlo sampling methods using {M}arkov chains and their
  applications.
\newblock {\em Biometrika\/}~{\em 57\/}(1), 97--109.

\bibitem[\protect\citeauthoryear{Jasra, Doucet, Stephens, and Holmes}{Jasra
  et~al.}{2008}]{jasra2008}
Jasra, A., A.~Doucet, D.~A. Stephens, and C.~C. Holmes (2008).
\newblock Interacting sequential {M}onte {C}arlo samplers for trans-dimensional
  simulation.
\newblock {\em Comput. Stat. Data Anal.\/}~{\em 52\/}(4), 1765--1791.

\bibitem[\protect\citeauthoryear{Jasra, Stephens, Doucet, and Tsagaris}{Jasra
  et~al.}{2008}]{jasra2008a}
Jasra, A., D.~A. Stephens, A.~Doucet, and T.~Tsagaris (2008).
\newblock Inference for {L}evy driven stochastic volatility models via adaptive
  {SMC}.
\newblock \verb=http://www.theodorostsagaris.com/svvg-DAS.pdf=.

\bibitem[\protect\citeauthoryear{Jasra, Stephens, and Holmes}{Jasra
  et~al.}{2007}]{jasra2007}
Jasra, A., D.~A. Stephens, and C.~C. Holmes (2007).
\newblock On population-based simulation for static inference.
\newblock {\em Statistics and Computing\/}~{\em 17\/}(3), 263--279.

\bibitem[\protect\citeauthoryear{Kirkpatrick, Gelatt, and Vecchi}{Kirkpatrick
  et~al.}{1983}]{kirkpatrick1983}
Kirkpatrick, S., C.~D. Gelatt, and M.~P. Vecchi (1983).
\newblock Optimization by simulated annealing.
\newblock {\em Science\/}~{\em 220\/}(4598), 671--680.

\bibitem[\protect\citeauthoryear{Kong, Liu, and Wong}{Kong
  et~al.}{1994}]{kong1994}
Kong, A., J.~S. Liu, and W.~H. Wong (1994).
\newblock Sequential imputations and {B}ayesian missing data problems.
\newblock {\em {Journal of the American Statistical Association}\/}~{\em
  89\/}(425), 278--288.

\bibitem[\protect\citeauthoryear{Liu and West}{Liu and
  West}{2001}]{SMCMPliuwest}
Liu, J. and M.~West (2001).
\newblock {\em Sequential {M}onte {C}arlo Methods in Practice}, Chapter 10:
  Combined Parameter and State Estimation in Simulation-Based Filtering.
\newblock Springer--Verlag New York.

\bibitem[\protect\citeauthoryear{Liu and Chen}{Liu and Chen}{1995}]{liu1995}
Liu, J.~S. and R.~Chen (1995).
\newblock Blind deconvolution via sequential imputations.
\newblock {\em {Journal of the American Statistical Association}\/}~{\em 90},
  567--576.

\bibitem[\protect\citeauthoryear{Liu and Chen}{Liu and
  Chen}{1998}]{liuchen1998}
Liu, J.~S. and R.~Chen (1998).
\newblock Sequential {M}onte {C}arlo methods for dynamic systems.
\newblock {\em Journal of the American Statistical Association\/}~{\em
  93\/}(443), 1032--1044.

\bibitem[\protect\citeauthoryear{Liu, Chen, and Wong}{Liu
  et~al.}{1998}]{liu1998}
Liu, J.~S., R.~Chen, and W.~H. Wong (1998).
\newblock Rejection control and sequential importance sampling.
\newblock {\em {Journal of the American Statistical Association}\/}~{\em
  93\/}(443), 1022--1031.

\bibitem[\protect\citeauthoryear{Metropolis, Rosenbluth, Rosenbluth, Teller,
  and Teller}{Metropolis et~al.}{1953}]{metropolis1953}
Metropolis, N., A.~W. Rosenbluth, M.~N. Rosenbluth, A.~H. Teller, and E.~Teller
  (1953).
\newblock Equation of state calculations by fast computing machines.
\newblock {\em The Journal of Chemical Physics\/}~{\em 21\/}(6), 1087--1092.

\bibitem[\protect\citeauthoryear{Neal}{Neal}{2001}]{neal2001}
Neal, R. (2001).
\newblock Annealed importance sampling.
\newblock {\em Statistics and Computing\/}~{\em 11\/}(2), 125--139.

\bibitem[\protect\citeauthoryear{Pasarica and Gelman}{Pasarica and
  Gelman}{2010}]{pasarica2010}
Pasarica, C. and A.~Gelman (2010).
\newblock Adaptively scaling the {M}etropolis algorithm using expected squared
  jumped distance.
\newblock {\em To appear: Statistica Sinica\/}.

\bibitem[\protect\citeauthoryear{Roberts and Rosenthal}{Roberts and
  Rosenthal}{2001}]{roberts2001}
Roberts, G. and J.~Rosenthal (2001).
\newblock Optimal scaling for various {M}etropolis-{H}astings algorithms.
\newblock {\em Statistical Science\/}~{\em 16\/}(4), 351--367.

\bibitem[\protect\citeauthoryear{Roberts and Rosenthal}{Roberts and
  Rosenthal}{2009}]{roberts2009}
Roberts, G.~O. and J.~S. Rosenthal (2009, June).
\newblock Examples of adaptive {MCMC}.
\newblock {\em Journal of Computational and Graphical Statistics\/}~{\em
  18\/}(2), 349--367.

\bibitem[\protect\citeauthoryear{Roberts and Tweedie}{Roberts and
  Tweedie}{1996}]{roberts1996}
Roberts, G.~O. and R.~L. Tweedie (1996).
\newblock Exponential convergence of {L}angevin distributions and their
  discrete approximations.
\newblock {\em Bernoulli\/}~{\em 2\/}(4), 341--363.

\bibitem[\protect\citeauthoryear{Sherlock and Roberts}{Sherlock and
  Roberts}{2009}]{sherlock2009}
Sherlock, C. and G.~Roberts (2009).
\newblock Optimal scaling of the random walk {M}etropolis on elliptically
  symmetric unimodal targets.
\newblock {\em Bernoulli\/}~{\em 15\/}(3), 774--798.

\bibitem[\protect\citeauthoryear{Stephens}{Stephens}{2000}]{stephens2000}
Stephens, M. (2000).
\newblock Dealing with label switching in mixture models.
\newblock {\em {Journal of the Royal Statistical Society, Series B}\/}~{\em
  62\/}(4), 795--809.

\bibitem[\protect\citeauthoryear{Storvik}{Storvik}{2002}]{storvik2002}
Storvik, G. (2002).
\newblock Particle filters for state-space models with the presence of unknown
  static parameters.
\newblock {\em IEEE Transaction on Signal Processing\/}~{\em 50}, 281--289.

\bibitem[\protect\citeauthoryear{{Ter Braak}}{{Ter Braak}}{2006}]{terbraak2006}
{Ter Braak}, C. J.~F. (2006).
\newblock A {M}arkov chain {M}onte {C}arlo version of the genetic algorithm
  differential evolution: easy {B}ayesian computing for real parameter spaces.
\newblock {\em Statistics and Computing\/}~{\em 16\/}(3), 239--249.

\bibitem[\protect\citeauthoryear{West}{West}{1993}]{west1993}
West, M. (1993).
\newblock Mixture models, {M}onte {C}arlo, {B}ayesian updating and dynamic
  models.
\newblock {\em Computing Science and Statistics\/}~{\em 24}, 325--333.

\bibitem[\protect\citeauthoryear{Whitley}{Whitley}{1994}]{whitley1994}
Whitley, D. (1994).
\newblock A genetic algorithm tutorial.
\newblock {\em Statistics and Computing\/}~{\em 4}, 65--85.

\end{thebibliography}

\end{document}